\documentclass[a4paper,10pt,twocolumn,superscriptaddress]{revtex4}
\usepackage{amsmath,amssymb}
\usepackage[final]{graphicx}

\begin{document}

\title{Doping of epitaxial graphene on SiC intercalated with hydrogen and its magneto-oscillations}
\author{Sergey Kopylov}
\affiliation{Department of Physics, Lancaster University, Lancaster, LA1~4YB, UK}
\author{Vladimir I. Fal'ko}
\affiliation{Department of Physics, Lancaster University, Lancaster, LA1~4YB, UK}
\author{Thomas Seyller}
\affiliation{Lehrstuhl f$\ddot{u}$r Technische Physik, Universit$\ddot{a}$t Erlangen-N$\ddot{u}$rnberg, 91056 Erlangen, Germany}

\begin{abstract}
We study the charge transfer between a quasi-free-standing monolayer graphene, produced by hydrogen
intercalation, and surface acceptor states. We consider two models of acceptor density of states
to explain the high hole densities observed in graphene and find the density responsivity to the gate voltage.
By studying magneto-oscillations of the carrier density we provide an experimental way to determine the
relevant model.
\end{abstract}

\maketitle

Among numerious ways of graphene fabrication \cite{Suspended,CVD,BN},
one of the promising methods for the top-down manufacturing of
electronic devices consists in the graphitization of Si-terminated surface of silicon carbide.
Graphene produced by such graphitization resides on an insulating
$(6\sqrt{3}\times 6\sqrt{3})$ R$30^\circ$ carbon buffer layer, which is easily spoiled
by vacancies and, typically, leads to high graphene doping \cite{ChargeBalance}.
Without special growth protocols, monolayer graphene on Si face of SiC appears to be highly n-doped,
with electron density at about $n_e=1\cdot 10^{13}$~cm$^{-2}$ level \cite{QHE1,Funct}.
Also, charged surface donors induce Coulomb scattering, which limits the mobility
of electrons in such a material. To improve the transport qualities
of graphene and decouple it from the substrate, it has been proposed to use
hydrogen intercalation \cite{Riedl}.
Hydrogen breaks Si-C bonds and converts the buffer layer into a quasi-free-standing monolayer graphene (QFMLG),
which is usually p-doped \cite{Speck} due to electron transfer from graphene
to acceptor states in the H-terminated surface of SiC.

\begin{figure}[tbp]
\centering
\includegraphics[width=0.9\columnwidth]{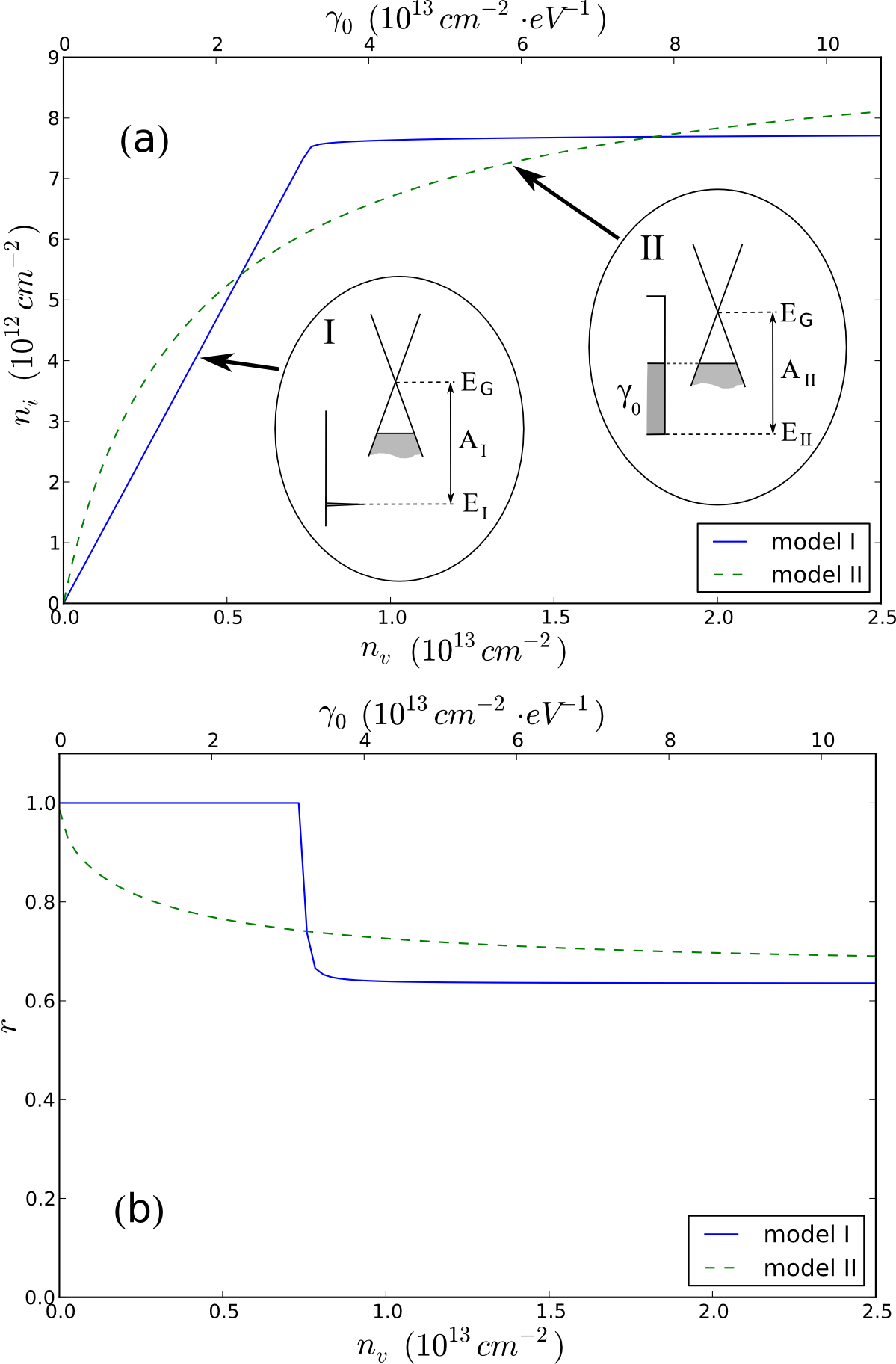}
\caption{a) The dependence of the hole density $n$ in QFMLG on the acceptor density (see $n_v$ axis for the model I
and $\gamma_0$ axis for the model II). The insets show charge distribution between graphene and acceptor states
for both models. b) Responsivity factor $r$ dependence on $n_v$.
The following parameters were used for both plots: $d=0.3$~nm, $A_I=0.6$~eV, $\Delta=5$~meV, $A_{II}=0.7$~eV.}
\label{fig:1}
\end{figure}

In this work we study the charge transfer between graphene and SiC in hydrogen-intercalated
epitaxial graphene, including the magneto-oscillations of the hole density in QFMLG.
We consider two limiting models for the charge transfer, which are different by the form of
the density of states of surface acceptors, $\gamma(\varepsilon)$.
In the model I we assume that all acceptor states are due to occasional unsaturated Si bonds:
vacancies in the hydrogen layer with the density $n_v$, energy $E_I$ and
narrow spectral density $\gamma_I(\varepsilon)=n_v
\text{exp}(-(\varepsilon-E_I)^2/\Delta^2)/(\sqrt{\pi}\Delta) \to n_v \delta(\varepsilon-E_I)$,
$\Delta \ll E_G-E_I$, where $E_G$ is the work function of graphene (Fig.~\ref{fig:1}, inset I).
In the model II we assume a uniform density of states of acceptor levels $\gamma(\varepsilon)=\gamma_0$
filled up to the energy $E_{II}$ (Fig.~\ref{fig:1}, inset II).
For both models we find the hole density in graphene and the responsivity factor describing
the effectiveness of QFMLG carrier density control by a gate voltage in a field-effect transistor.
We also analyse the oscillations in the carrier density dependence on a magnetic field,
and show that this can be used to distinguish experimentally between the two limiting doping models.

The electron transfer from QFMLG to surface acceptor states is described by the following equation,
\begin{equation}
n-n_g = \int\limits^{\varepsilon(n)}_{E_{min}} d\varepsilon\, \gamma(\varepsilon); \quad
\varepsilon(n) = E_G - \frac{e^2 d}{\varepsilon_0} (n-n_g) + \varepsilon_F(n),
\label{eq:balance}
\end{equation}
where $n$ is the density of holes in graphene, $n_g=CV_g/e$ ($V_g$ is gate voltage),
$d$ is the distance between SiC and QFMLG,
the Fermi energy (relative to the graphene Dirac point) is $\varepsilon_F(n)=-\hbar v \sqrt{\pi n}$,
and $E_{min}=-\infty$ for the model I and $E_{min}=E_{II}$ for the model II.
Note that integral in Eq.~\ref{eq:balance} is taken over the electron (rather than hole) energy.
Each of the models is characterized by one energy parameter: $A=E_G-E_I$ for model I and
$A=E_G-E_{II}$ for model II.

The value of the carrier density for non-gated structures can be obtained by solving Eq.~(\ref{eq:balance}) for $n_g=0$.
In Fig.~\ref{fig:1}a we illustrate the hole density dependence in graphene on the amount of acceptors on 
hydrogenated SiC. For the model I, it is
\begin{equation}
n_I = \text{min}\left[n_v, n_b \right]; \quad
n_b=\frac{4A^2}{\left( \hbar v\sqrt{\pi} + \sqrt{\hbar^2 v^2\pi + \frac{4e^2 d A}{\varepsilon_0}} \right)^2}.
\end{equation}
At a small $n_v$, all acceptor levels are occupied and $n_I=n_v$.
As acceptor density $n_v$ increases, graphene doping saturates at $n_I\approx 1\cdot 10^{13}$~cm$^{-2}$.
Model II also shows a saturation of the carrier density, but with a smoother crossover,
\begin{equation}
n_{II} = \frac{4A^2}{\left( \hbar v\sqrt{\pi} + \sqrt{\hbar^2 v^2\pi +
4A\left(\frac{1}{\gamma_0} + \frac{e^2 d}{\varepsilon_0}\right)} \right)^2}.
\end{equation}

\begin{figure}[tbp]
\centering
\includegraphics[width=0.97\columnwidth]{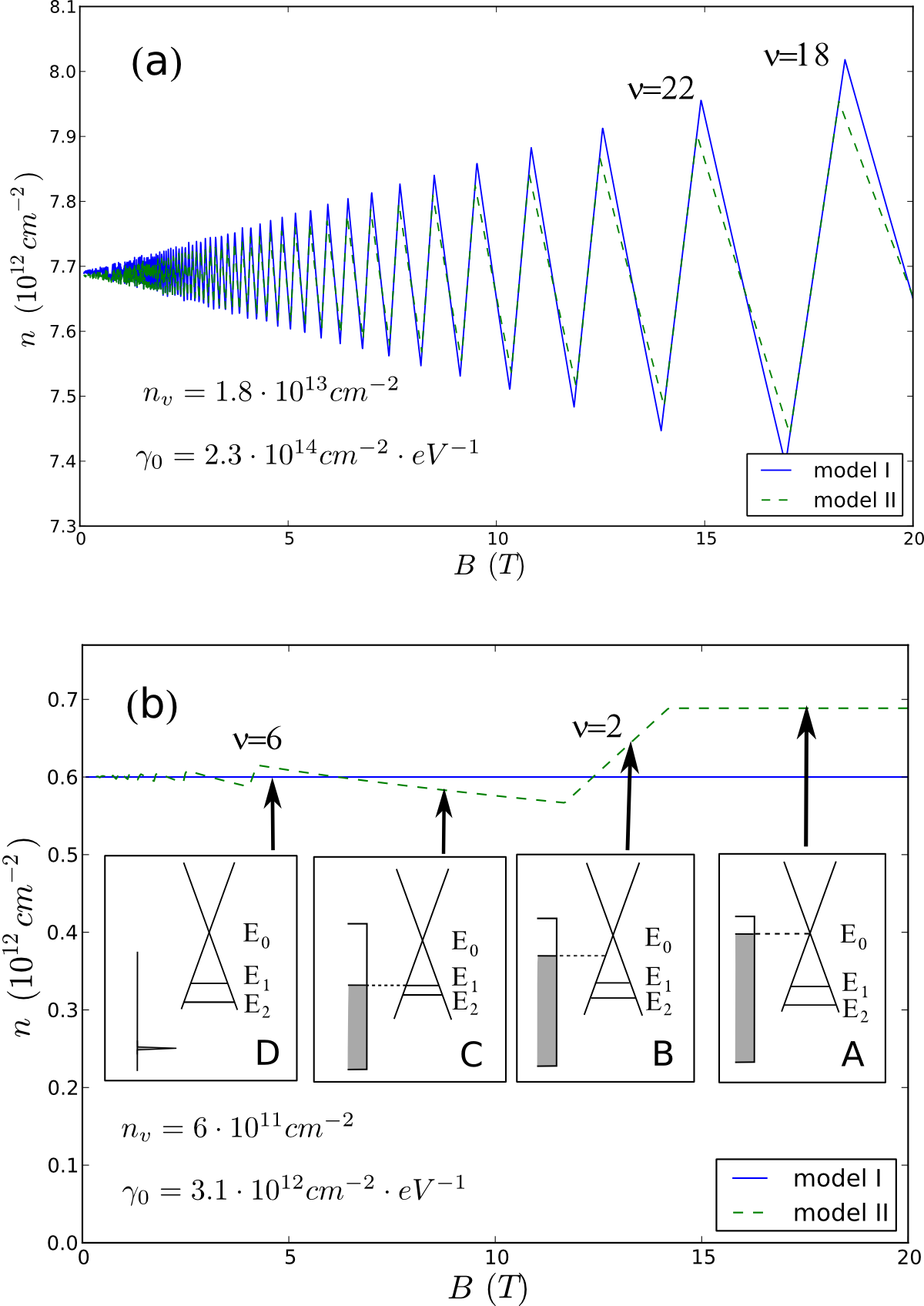}
\caption{a) The magneto-oscillations of the hole density $n$ in QFMLG in the regimes of large acceptor density.
b) $n(B)$ dependence at small acceptor density. The energy structure in different regimes is shown
in the insets A-C (model II) and inset D (model I). The parameters used here were chosen for illistration purposes only.}
\label{fig:2}
\end{figure}

The effectiveness of using QFMLG transistors can be characterized by the responsivity factor, $r=dn/dn_g$.
Then, $r=1$ corresponds to the regime of the effective transistor operation, while $r\ll 1$ indicates that it is difficult
to change carrier density in graphene. The responsivity factors for the models I and II,
\begin{equation}
r_I = 1 - \frac{1+\text{sign}(n_v-n_b)}{2\sqrt{1 + 4e^2dA/(\pi\hbar^2 v^2\varepsilon_0)}};
\nonumber
\end{equation}
\begin{equation}
r_{II} = 1 - \frac{1}{\sqrt{1 + \frac{4A}{\pi\hbar^2 v^2}\left(\frac{1}{\gamma_0}+\frac{e^2d}{\varepsilon_0}\right)}},
\nonumber
\end{equation}
are compared graphically in Fig.~\ref{fig:1}b.

Due to the charge transfer between graphene and acceptors in the hydrogen layer on SiC,
carrier density of QFMLG change upon the variation of the magnetic field $B$.
For high acceptor densities ($n_v\gg 10^{12}$~cm$^{-2}$, $\gamma_0\gg 10^{13}$~cm$^{-2}$eV$^{-1}$),
Fig.~\ref{fig:2}a shows the oscillations in $n(B)$ dependence,
which appear due to a non-zero density of acceptor states at the Fermi level.
Similar oscillations were found in graphene on a non-hydrogenated SiC surface
\cite{QHE}, and used to explain
the presence of a wide $\nu=2$ plateau in QHE. Both models give close results for the $n(B)$ dependence,
which reveals two different regimes. (a) The regions with a negative slope
indicating the pinning of the Fermi level to one of the Landau levels in graphene, $E_N=-\sqrt{2e\hbar v^2 B N}$.
To evaluate $n(B)$ dependence in this regime we solved Eq.~(\ref{eq:balance}),
with $\varepsilon_F=E_N$. (b) The regions with a positive slope corresponding to the pinning of filling factor
$\nu=nh/(eB)=4N+2$, for which the Fermi level in the system lies in the acceptor band, between Landau levels.

The difference between the density oscillations in graphene expected on the basis of the models I and II
is quite pronounced in structures with low acceptor densities
($n_v\lesssim 10^{12}$~cm$^{-2}$, $\gamma_0\lesssim 10^{13}$~cm$^{-2}$eV$^{-1}$),
as shown on Fig.~\ref{fig:2}b. In this regime no oscillations are observed for model I,
in contrast to the model II, where pinning of $\nu=2$ takes place over magnetic field
interval of several Tesla. In the latter case, one can identify the following charge transfer regimes
illustrated using sketches in Fig.~\ref{fig:2}b.
At high magnetic fields ($B>B_1=hA/2e/(\gamma^{-1}_0+e^2d/\varepsilon_0)$) the carrier density
remains constant, since the 0th Landau level (LL) is partially occupied and the Fermi
level in graphene is pinned to the Dirac point (inset A in Fig.~\ref{fig:2}b). At lower magnetic fields
($2A^2/(\sqrt{e\hbar v^2}+\sqrt{e\hbar v^2 + 4eA/(\gamma^{-1}_0+e^2d/\varepsilon_0)/h})^2<B<B_1$)
0th LL becomes completely unoccupied and the Fermi level
sticks to acceptor levels between 0th and 1st LLs (Fig.~\ref{fig:2}b, inset B). This regime, observed within
a window of several T, is responsible for widening the $\nu=2$ QHE plateau.
The region C in Fig.~\ref{fig:2}b refers to the pinning of the Fermi level to the 1st LL $E_1$.

In conclusion, we have considered two models of the charge transfer between hydrogen intercalated graphene
and acceptors in SiC surface. The first model describes the case when all acceptor states are created
by unsaturated Si bonds with energies from a narrow window. The opposite limit of a wide acceptor
energy distribution is covered by the second model. Both models predict a saturation behavior
of carrier density in graphene as the acceptor density increases, however, the magnetic field
dependence of charge transfer in them is very different, especially in structures with a low initial doping
$n\lesssim 10^{12}$~cm$^{-2}$ by holes. As shown in Fig.~\ref{fig:2}b,
the model II features a wide (several T) $\nu=2$ QHE plateau, while the model I reveals no
$n(B)$ dependence. One of the ways to detect pinning of filling factor $\nu=2$ is to measure
the activation energy $T^*$ and/or breakdown current of QHE \cite{QHE}.
The width of the peak in the dependence of $T^*$ and breakdown current on the magnetic field
in the vicinity of $\nu=2$ QHE plateau determines the region of filling factor pinning.
Alternatively, one can determine the filling factor by looking at the Landau level occupancy
using scanning tunneling spectroscopy as in Ref.~\cite{STS}.

\end{document}